\documentstyle[aps]{revtex}

\begin{document}
\author{Hua L\"{u}}
\title{N-Partner Secure Direct Communication Based on Quantum Nonlocality}
\address{Department of mathematics and physics, Hubei polytechnic university, Wuhan,\\
430068, China}
\author{Qing-Yu Cai}
\address{Wuhan Institute of Physics and Mathematics, The Chinese Academy of Sciences,%
\\
Wuhan 430071, China}
\maketitle

\begin{abstract}
A multipartner secure direct communication protocol is presented, using
quantum nonlocality. Security of this protocol is based on `High fidelity
implies low entropy'. When the entanglement was successfully distributed,
anyone of the multipartner can send message secretly by using local
operation and reliable public channel. Since message transfered only by
using local operation and public channel after entanglement successfully
distributed, so this protocol can protect the communication against
destroying-travel-qubit-type attack.
\end{abstract}


Since the pioneering work of Bennett and Brassard published in 1984 [1],
different quantum key distribution protocols has been presented [2-15].
Different from the key distribution protocol, some quantum direct secure
communication (QDSC) protocols have been shown recently [16-18], which
permits important message can be communication directly without first
establishing a random key to encrypt them. However, these protocols [1-18]
only permit message transferred from the sender (Alice) to the receiver
(Bob). In reference [19], a protocol was presented that allows Alice and Bob
communicate with each other coequally. When Alice want to secretly send
message to two or more persons at the same time, the protocols based on
entanglement of two qubit are feeble. In this paper, we show a secure direct
communication protocol that allows multipartner instantaneously
communicating secretly by using .

$The$ $multipartner$ $protocol$. Three qubit can be entangled in a
Greenberger-Horne-Zeilinger (GHZ) state $|\phi _{0}>$ [20] 
\begin{equation}
|\phi _{0}>=\frac{1}{\sqrt{2}}(|0>|0>|0>+|1>|1>|1>),
\end{equation}
where $|0>$ and $|1>$ are the up and down eigenstate of the $\sigma _{z}$,
the photon polarization operator. Suppose three partners, Alice , Bob and
Charlie, want to secretly communicate together. First of all, Alice prepares
three qubits in a GHZ state $|\phi _{0}>$. She keeps one qubit and sends
each of the other two qubits to Bob and Charlie. When Bob and Charlie
receive the travel, both of them send a classical signal as receipt to Alice
through public channel. When Alice receives their receipt, she states this
is a message mode in public. Then Alice, bob and Charlie perform a
measurement in basis $B_{z}=\{|0>,|1>\}$. If Alice's measurement is $|0>$ ($%
|1>$), then she knows Bob and Charlie's measurement outcome are $|0>$ ($|1>$%
). When she wants to send a classical bit `0' to Alice and Bob, if her
measurement is $|0>$, then she `say yes' in public. If her measurement is $%
|1>$, then she `say no' through public channel to both Bob and Charlie. When
Alice wants to send a logical `1' to Bob and Charlie, if her measurement
outcome is 
\mbox{$\vert$}%
0%
\mbox{$>$}%
, she `say no' in public. Else, she 'say yes' to Bob and Charlie through
public channel. By default, Alice, Bob and Charlie are in message mode and
communicate the way described above. With probability $c$, Alice switches to
control mode. In control mode. when Alice receives Bob and Charlie's
receipt, she performs a measurement randomly in basis $B_{z}$ or $%
B_{x}=\{|+>,|->\}$, where $|+>=\frac{1}{\sqrt{2}}(|0>+|1>)$, $|->=\frac{1}{%
\sqrt{2}}(|0>-|1>)$. Then she announces her measurement outcome. Bob and
Charlie also switch to control mode. They perform measurement in the same
basis Alice used. If their measurement outcomes do not coincide when both of
them use the basis $B_{z}$, there is an eavesdropper-Eve in linear. When
they use the basis $B_{x}$, their measurement outcome should be in the
states $|\pm >$. Using 
\begin{equation}
\sigma _{x}|+>=|+>,\sigma _{x}|->=-|->,
\end{equation}
when the products of $\sigma _{x}^{A}$ (Alice), $\sigma _{x}^{B}$ (Bob), and 
$\sigma _{x}^{C}$ (Charlie) is $-1$, there is an Eve in line. The
communication stops. Else, this communication continues.

$Security$ $proof$. To effectively eavesdrop Alice's information, Eve has to
distinguish what state Bob or Charlie obtains since she has no access to
Alice's qubit. It has been presented that there two inequivalent forms of
three qubits entangled state [21]. One is namely GHZ state. The other is $%
|W> $ 
\begin{equation}
|W>=\frac{1}{\sqrt{3}}(|1>|0>|0>+|0>|1>|0>+|0>|0>|1>).
\end{equation}
These two different states can not be converted by means of SLOCC [21].
After Eve's attack, the state will become a product state or still is a
GHZ-type state 
\begin{eqnarray}
|\phi _{1} &>&=\frac{1}{\sqrt{2}}(|0>|0>|0>-|1>|1>|1>), \\
|\phi _{2} &>&=\frac{1}{\sqrt{2}}(|1>|0>|0>+|0>|1>|1>), \\
|\phi _{3} &>&=\frac{1}{\sqrt{2}}(|1>|0>|0>-|0>|1>|1>), \\
|\phi _{4} &>&=\frac{1}{\sqrt{2}}(|0>|1>|0>+|1>|0>|1>), \\
|\phi _{5} &>&=\frac{1}{\sqrt{2}}(|0>|1>|0>-|1>|0>|1>), \\
|\phi _{6} &>&=\frac{1}{\sqrt{2}}(|0>|0>|1>+|1>|1>|0>), \\
|\phi _{7} &>&=\frac{1}{\sqrt{2}}(|0>|0>|1>-|1>|1>|0>).
\end{eqnarray}
Together with $|\phi _{0}>$, these eight states compose a complete base.
Suppose the state $|\phi _{0}>$ becomes $\rho $ under Eve's attack. $\rho $
must be a linear combination of $|\phi _{i}>$. The fidelity [22] of $\rho $
and $|\phi _{0}>$ is 
\begin{eqnarray}
F(|\phi _{0} &>&,\rho )=tr\sqrt{<\phi _{0}|\rho |\phi _{0}>|\psi ^{-}><\psi
^{-}|} \\
&=&\sqrt{<\phi _{0}|\rho |\phi _{0}>}  \nonumber
\end{eqnarray}
`High fidelity implies low entropy.' Assume $F(|\phi _{0}>,\rho
)^{2}=1-\gamma $. The information from $\rho $ is bounded by the Holevo
quantity, $\chi (\rho )$ [23]. From 
\begin{equation}
\chi (\rho )=S(\rho )-\sum_{i}p_{i}\rho _{i},
\end{equation}
we know that $S(\rho )$ is the upper bound of the information Eve can gain.
The entropy of $\rho $ is bounded above by the entropy a diagonal density
matrix $\rho _{\max }$ with diagonal entries $1-\gamma $, $\frac{\gamma }{7}%
,...,\frac{\gamma }{7}$. The entropy of $\rho _{\max }$ is 
\begin{equation}
S(\rho _{\max })=-(1-\gamma )\log (1-\gamma )-\gamma \log \frac{\gamma }{7}.
\end{equation}
Let us consider the relation between the information Eve can gain and
detection probability $d$. When $\rho $ is anyone of $|\phi _{i}>$, $%
i=2,...,7$, Alice, Bob and Charlie can perform the measurement in basis $%
B_{z}$ and public their measurement outcomes. When their measurement
outcomes do not coincide, there is an Eve in line. When $\rho $ is the state 
$|\phi _{1}>$, Alice, Bob and Charlie can perform a measurement in basis $%
B_{x}$ and public their measurement outcome$:$%
\[
\begin{array}{llll}
A & B & C & product \\ 
+1 & +1 & -1 & -1 \\ 
+1 & -1 & +1 & -1 \\ 
-1 & +1 & +1 & -1 \\ 
-1 & -1 & -1 & -1
\end{array}
\]
comparing with the measurement outcomes of $|\phi _{0}>$: 
\[
\begin{array}{llll}
A & B & C & product \\ 
+1 & +1 & +1 & +1 \\ 
+1 & -1 & -1 & +1 \\ 
-1 & +1 & -1 & +1 \\ 
-1 & -1 & +1 & +1
\end{array}
. 
\]
When they find out that the product of their outcomes is $-1$, there is an
Eve in line. Since the fidelity $F(|\phi _{0}>,\rho )^{2}=1-\gamma $, then
the detection probability $d$ is approximately linearly dependent on the
quantity $d\geq \gamma /2$. When $\gamma =0$, there is $S(\rho _{\max })=0$,
i.e., Eve can get no information from the communication. When $S(\rho _{\max
})>0$, i.e., Eve can get some information from the communication, she has to
face a nonzero risk to be detected.

This protocol is quasisecure for a direct communication but is secure for a
quantum key distribution protocol. Also, multipartner can communication
using two-particle entanglement network. However, our protocol can provide a
secure direct communication between three partners. Actually, Alice, Bob and
Charlie are coequal in this protocol. So they can communicate with others
coequally when the entanglement was successfully shared. Consider there are
N partner want to communicate securely. They can share N- particle GHZ
states. With local measurement and classical message, they can communicate
securely. When N-particle entanglement was successfully shared, no qubit has
to be exchanged in quantum channel. Then the destroying-travel-qubit-type
attack [19] is invalid.

Entanglement of three photons has been demonstrate experimental [24]. And
the quantum nonlocality of four photons entanglement has already been
reported [25]. The experimental feasibility of our protocol, can be realized
with today's technology.

\section{references.}

[1] C. H. Bennett, and G. Brassard, in Proceedings of the IEEE international
Conference on Computers, Systems and Signal Processing, Bangalore, India
(IEEE, New York, 1984), pp. 175-179.

[2] A. K. Ekert, Phys. Rev. Lett. 67, 661 (1991).

[3] C. H. Bennett, G. Brassard, and N. D. Mermin, Phys. Rev. Lett. 68, 557
(1992).

[4] C. H. Bennett, Phys. Rev. Lett. 68, 3121 (1992).

[5] C. H. Bennett and S. J. Winsner, Phys. Rev. Lett. 69, 2881 (1992).

[6] L. Goldenberg and L. Vaidman, Phys. Rev. Lett. 75, 1239 (1995).

[7] B. Huttner, N. Imoto, N. Gisin, and T. Mor, Phys. Rev. A 51, 1863 (1995).

[8] M. Koashi and N. Imoto, Phys. Rev. Lett. 79, 2383 (1997).

[9] D. Bruss, Phys. Rev. Lett. 81, 3018 (1998).

[10] W. Y. Hwang, I. G. Koh, and Y. D. Han, Phys. Lett. A 244, 489 (1998).

[11] H.-K. Lo and H. F. Chau, Science 283, 2050 (1999).

[12] A. Cabell, Phys. Rev. Lett. 85, 5635 (2000).

[13] A. Beige, B.-G. Englert, C. Kurtsiefer, and H. Weinfurter, Acta. Phys.
Pol. A 101, 357 (2002).

[14] B. S. Shi, Y. K. Jiang, and G. C. Guo, Appl. Phys. B: Laser Opt. B70,
415(2002).

[15] P. Xue, C. F. Li, and G. C. Guo, Phys. Rev. A 65, 022317 (2002).

[16] A. Beige et al., Acta Phys. Pol. A 101, 357 (2002).

[17] K. Bostroem and T. Felbinger, Phys. Rev. Lett. 89, 187902 (2002).

[18] Fu-Guo Deng, Gui Lu Long, and Xiao-Shu Liu, Phys. Rev. A 68, 042317
(2003).

[19] Qing-yu Cai, arXiv:quant-ph/0309108 (unpublished).

[20] D. M. Greenberger, M. A. Horne, and A. Zeilinger, in Bell's Theorem,
Quantum Theory, and Conceptions of the Universe, edited by M. Kafatos
(Kluer, Dordrecht, 1989), p. 69.

[21] W. Duer, G. Vidal, and J. I. Cirac, Phys. Rev. A 62, 062314 (2000).

[22] C. A. Fuchs, arXiv: quant-ph/9601020; H. Barnum, C. M. Caves, C. A.
Fuchs, R. Jozsa, and B. Schumacher, Phys. Rev. Lett. 76, 2818 (1996).

[23] M. A. Nielsen and I. L. Chuang, Quantum computation and Quantum
Information (Cambridge University Press, Cambridge, UK, 2000).

[24] D. Bouwmeester et al., Phys. Rev. Lett. 82, 1345 (1999).

[25] J.-W. Pan et al., Nature (London) 403, 515 (2000).

\end{document}